\documentclass[journal]{IEEEtran}

\usepackage{times}
\usepackage{epsfig}
\usepackage{graphicx}
\usepackage{amsmath}
\usepackage{amssymb}
\usepackage{colortbl}
\usepackage{multirow}
\usepackage{soul}
\usepackage{adjustbox}
\usepackage{array}
\usepackage{setspace}
\usepackage{lscape}
\usepackage{mathtools}
\usepackage{hyperref}
\usepackage{pgfplots}
\pgfplotsset{compat=newest}

\usepackage{enumitem}
\usepackage{multirow}
\usepackage{rotating}
\usepackage{caption}
\usepackage{cite}
\usepackage{amsmath,amssymb,amsfonts}
\usepackage{algorithmic}
\usepackage{graphicx}
\usepackage{textcomp}
\usepackage{caption} 
\usepackage{subcaption}
\usepackage{amssymb}
\usepackage{pifont}
\newcommand{\cmark}{\ding{51}}%
\newcommand{\cmarkbig}{\ding{52}}%
\usepackage{fixfoot,xspace}

\begin{document}

\DeclareFixedFootnote*{\dbfootnote}{\url{https://github.com/BiDAlab/ChildCIdb_v1}}
\DeclareFixedFootnote*{\gafootnote}{\url{https://github.com/BiDAlab/GeneticAlgorithm}}

\title{ChildCI Framework: Analysis of Motor and Cognitive Development in Children-Computer Interaction for Age Detection}

\author{
Juan Carlos Ruiz-Garcia, Ruben Tolosana, Ruben Vera-Rodriguez, \\ Julian Fierrez, Jaime Herreros-Rodriguez \\

\thanks{J.C. Ruiz-Garcia, R. Tolosana, R. Vera-Rodriguez and J. Fierrez are with the Biometrics and Data Pattern Analytics - BiDA Lab, Escuela Politecnica Superior, Universidad Autonoma de Madrid, 28049 Madrid, Spain (e-mail: juanc.ruiz@uam.es, ruben.tolosana@uam.es; ruben.vera@uam.es; julian.fierrez@uam.es).

J. Herreros-Rodriguez is with the Hospital Universitario Infanta Leonor, 28031 Madrid, Spain (e-mail: hrinvest@hotmail.com).}}

\maketitle

\begin{abstract}
    This article presents a comprehensive analysis of the different tests proposed in the recent ChildCI framework\dbfootnote, proving its potential for generating a better understanding of children's neuromotor and cognitive development along time, as well as their possible application in other research areas such as e-Health and e-Learning. In particular, we propose a set of over 100 global features related to motor and cognitive aspects of the children interaction with mobile devices, some of them collected and adapted from the literature. 

    Furthermore, we analyse the robustness and discriminative power of the proposed feature set including experimental results for the task of children age group detection based on their motor and cognitive behaviours. Two different scenarios are considered in this study: \textit{i)} single-test scenario, and \textit{ii)} multiple-test scenario. Results over 93\% accuracy are achieved using the publicly available ChildCIdb\_v1 database (over 400 children from 18 months to 8 years old), proving the high correlation of children's age with the way they interact with mobile devices.
\end{abstract}


\section{Introduction}\label{sec1:intro}

\IEEEPARstart{T}{echnology} has become a very important aspect of our lives in recent decades. In particular, mobile devices play an essential role in our daily basis (e.g., work, relationships, communications, business, etc.). This also affects children, who are exposed to these devices from an early age~\cite{Antle2021}. Recent studies corroborate this fact~\cite{Kabali2015, kilicc2019exposure}. For example, Kabali \textit{et al.} conducted a study in~\cite{Kabali2015} with 350 children aged 6 months to 4 years concluding that 96.6\% of children use mobile devices, and most started using them before the age of 1 year. In addition, around 75\% of children by the age of 4 years already have their own mobile device. Similar conclusions were obtained in~\cite{kilicc2019exposure}, where 422 parents of children aged from birth to 5 years were interviewed and 75.6\% of them indicated that their children had already used mobile devices at that age. Moreover, and due to the global pandemic of COVID-19 since 2020, the use of mobile devices has been rapidly increased as preschools, kindergartens, and schools were closed down for several months in most countries around the world. As a result, traditional face-to-face education was replaced to virtual learning environments (e-Learning)~\cite{antle2020child}.

Despite the high technological evolution and the application of it in children scenarios, the assessment of the correct motor and cognitive development of children is still evaluated using traditional approaches that are manual, time-consuming, and provide qualitative results that are difficult to interpret. This is one of the main motivations of our ChildCI framework~\cite{Tolosana2021}: the proposal of automatic methods that quantify the motor and cognitive development of the children through the interaction with mobile devices, using both the stylus and the finger/touch. As a first step towards that future goal, in this article we first evaluate the discriminative power of the tests proposed in the ChildCI framework, trying to shed some light on the following questions: Is there any relationship between children's chronological age and their motor and cognitive development when interacting with the tests proposed in ChildCI framework? Is there any relationship between the age, the type of test, and writing input (stylus/finger) considered? The answers to these questions could provide very interesting insights for the research community and the proposal of automatic and usable methods to better quantify the development of the children.

The main contributions of the present work are:

\begin{itemize}
    \item An in-depth revision of recent works studying children's interactions on mobile devices (using both finger and stylus), as well as the analysis of motor and cognitive development.
    \item Validate the potential of the different tests included in the ChildCI framework in terms of the motor and cognitive development of the children. We propose a feature set with over 100 global features based on cognitive and motor aspects of children while interacting with mobile devices, some of them collected and adapted from the literature.
    \item Analyse whether there is any relationship between the chronological age of the children and their motor and cognitive development while interacting with the tests included in ChildCI. To shed some light on this, experiments are carried out for the task of children age group detection based on their motor and cognitive behaviors. Three groups are considered: 1 to 3 years, 3 to 6 years, and 6 to 8 years. Different acquisition inputs are considered in the analysis, i.e., stylus and finger. In addition, single- and multiple-test experiments are studied. Experiments are performed including several automatic feature selection techniques and machine learning approaches.
\end{itemize}

The remainder of the article is organised as follows. Sec.~\ref{sec2:related_works} summarises previous studies on children's touch and stylus interaction, as well as some sets of features successfully used in literature for different lines of work. In the following, we describe in Sec.~\ref{sec3:childci_framework} the ChildCI framework and database used in the experimental work carried out. Sec.~\ref{sec4:method} describes the proposed feature set based on motor and cognitive aspects of the children. Sec.~\ref{sec5:experiments} describes the experimental protocol and the results achieved for children age group detection task. Finally, Sec.~\ref{sec6:conclusion_future_work} draws the final conclusions and future work.

\section{Related Works}\label{sec2:related_works}

\subsection{Children Interaction: Stylus vs Finger}\label{subsec21:stylus_finger}

From such an early age and throughout their development, children experience different evolutionary stages in which their physiological and cognitive capacities improve through continuous interaction with the world they live. Piaget and Inhelder were the leaders of the study of children's motor and cognitive development and, according to their theory~\cite{piaget2008}, children pass in a fixed sequence through four universal stages of development: \textit{i) Sensorimotor} (from birth to 2 years), children focus on acquiring knowledge by using their senses to touch, smell, see, taste, and hear the objects around them; \textit{ii) Preoperational} (2-7 years), their language and thinking improve together with their motor skills. In addition, at this age children are egocentric in their thinking and it is still difficult for them to empathise with other people's feelings; \textit{iii) Concrete Operational} (7-11 years), children begin to use more logical thinking to solve problems, starting to improve their empathic abilities significantly; and \textit{iv) Formal Operational} (11 years to adulthood), they gain the ability to use abstract cognitive functions to think more about moral, philosophical, ethical, social, and political issues.

Children's interaction with mobile devices has been evaluated and analysed by multiple research studies in recent decades. Focusing on the first stage of Piaget's theory (Sensoriomotor, 0-2 years) there is not much work on the interaction analysis of children under the age of 2 with touchscreen devices, mainly due to the difficulty of capturing data with children at that age. If we focus on touch mobile interactions, Morante \textit{et al.} presented a very interesting article in this line in~\cite{Morante2016}. In that work, the authors analysed the behaviors of children aged from 0 to 2 years. They concluded that children at 1 year of age can use the tap gesture intentionally to perform actions and at 2 years they are already able to understand some gestures such as tap and drag to navigate through apps. In \cite{Hourcade2015} the authors assessed the mobile interaction of children aged 1 to 2 years through the analysis of videos from YouTube while they were recorded interacting with mobile devices. They concluded that children under 17 months tend to use both hands for interaction, an aspect that decreases sharply with age, leading to single-hand use.

Looking at the second stage of Piaget's theory (Preoperational, 2-7 years), several studies have analysed the children's interaction with mobile devices, in contrast with the first stage. For example, in the work presented by Vatavu \textit{et al.} \cite{vatavu2015touch}, a database of 89 children aged 3 to 6 years and 30 young adults was presented. This database was also considered in the experimental protocol of Vera-Rodriguez \textit{et al.}~\cite{Vera-Rodriguez2020}. In that work, classification rates above 96\% were achieved for the adult-child detection task using an automatic system based on neuromotor skills. A similar research line was studied by Nacher \textit{et al.} in~\cite{nacher2015multi}, where the authors proposed a set of 8 different tests on a mobile device in order to measure the ability of children aged 2-3 years to perform touch gestures. The results showed that simple gestures such as tap, drag, and one-finger rotation can be performed by children in most cases. However, performing more complex gestures such as double tap, scale down, long press, and two-finger rotation is strongly influenced by the age of the child, with the older children's group performing them easier and quicker than the younger ones. Similar conclusions were obtained in \cite{chen2020examining}. The authors found different children's interaction behaviors with mobile devices by analysing the correlation between factors such as their age, grade level, motor and cognitive development, and how they performed touchscreen interaction tasks (target acquisition and gesture detection).

Interaction with mobile devices is not only done through the use of the finger, but also through an stylus~\cite{Tolosana2021DeepWrite, Tolosana2022SVC}. In general, writing and drawing require greater motor and cognitive development than simple touch gestures. Children start scribbling around the age of 2 years~\cite{Price2015}. In~\cite{remi2015exploring}, Rémi \textit{et al.} studied the way children aged 3-6 years perform scribbling activities, concluding that there are significant differences in motor skills depending on the age. Another interesting work in this line is presented in \cite{tabatabaey2015analyses} considering children 6-7 years old. The authors analysed the correlation between the performance of polygonal shape drawing and levels in handwriting performance. The results proved that there are different children's drawing strategies that differ in their writing performance.

\subsection{Motor and Cognitive Development}\label{subsec22:related_features}

The way certain actions and gestures are performed on a touchscreen device determines the behavioral patterns associated with a specific individual~\cite{Singh2023, Schadenberg2022}. In particular, the correct analysis and quantification of the motor and cognitive development of the children is based on a good definition of robust and discriminative features for the task. Previous studies in the field of Human-Computer Interaction (HCI) could provide interesting features that, after adapting them, could be very useful to analyse motor and cognitive aspects of the children.

For example, in~\cite{Ishii2020} Ishii \textit{et al.} developed a simple quantitative method to diagnose tremor using hand-drawn spirals and artificial intelligence. The Archimedes spiral is the reference test for the clinical diagnosis of diseases such as essential tremor or Parkinson. In that study, patients used a stylus to trace a spiral on a printed reference spiral and, by comparing the lengths of the reference spiral and the traced one, the total area of deviation between both was calculated, achieving results with success rates up to 79\% in detecting people with essential tremor. In a similar work~\cite{Sole-Casals2019}, Solé-Casals \textit{et al.} proposed a new set of 34 features using only the \textit{x} and \textit{y} coordinate points of the strokes made by patients as they traced the Archimedes spiral using a pen stylus on a graphics tablet. In addition to tremor assessment, in~\cite{Lin2018} the authors proposed a test paradigm on a graphic tablet using different parameters to automatically quantify tremor characteristics and severity in real-time by extracting three parameters: \textit{i)} the mean radial difference per radian, \textit{ii)} the mean radial difference per second, and \textit{iii)} the area under the curve of the frequency spectrum for the velocity. Tremor is directly related to fine motor actions such as pinching, writing, drawing and other small movements. Therefore, it is interesting to analyse the level of tremor in children as they grow up, because it will be higher or lower depending on their motor skills development.

An interesting article in this line was the work presented by Xu \textit{et al.} in~\cite{Hui2014}, where a variety of touch gestures were used to enhance the security and privacy of users based on the touch operations performed on their smartphone screens. Through the analysis of touch gestures such as swipe, drag and drop, tap or pinch, among others, the authors proposed a total of 132 features that identify the way each user interacts with the mobile device, achieving an Equal Error Rate (EER) of around 10\% for all types of gestures and 1\% for the swipe operation where the Largest Deviation Point (LDP) was considered. Another interesting study on this line was carried out by Vatavu \textit{et al.} in~\cite{vatavu2015child}, where through the features extracted using the touch coordinates \textit{x} and \textit{y}, it was possible to detect the age group of the users reaching up to 86.5\% accuracy. A similar study was conducted in~\cite{Zaccagnino2021}, where the authors proposed a novel approach to protect society from online threats through the interaction of 147 participants with six micro-games in an Android app. A dataset of more than 9,000 touch gestures was created, characterising how participants interact with the device and achieving results up to 88\% accuracy detecting impostors.

\section{ChildCI Framework}\label{sec3:childci_framework}

As we preliminary presented in~\cite{Tolosana2021}, ChildCI is an on-going project mainly intended to improve the understanding of children's motor and cognitive development along time through the interaction with mobile devices. Stylus and finger are used as acquisition tools, capturing data and storing it in our novel ChildCI database (ChildCIdb\_v1\dbfootnote). This is a database collected in collaboration with the school GSD Las Suertes in Madrid (Spain), which is planned to be extended yearly, allowing for interesting longitudinal studies. To the best of our knowledge, ChildCIdb\_v1 is the largest and most diverse publicly available Child-Computer Interaction (CCI) dataset to date on the topic of the interaction of children with mobile devices. It is composed of 438 children in the ages from 18 months to 8 years, grouped in 8 different educational levels according to the Spanish education system. In addition, during the capture process other interesting information from the children is also collected: \textit{i)} previous experience using mobile devices, \textit{ii)} grades at the school, \textit{iii)} attention-deficit/hyperactivity disorder (ADHD), \textit{iv)} birthday date, \textit{v)} prematurity (under 37 weeks gestation). All this additional metadata makes the project more powerful and interesting, allowing for multiple lines of future research. This dataset is considered in the experimental framework of this study.

\begin{figure*}[!]
\begin{center}
   \includegraphics[width=\linewidth,height=\textheight]{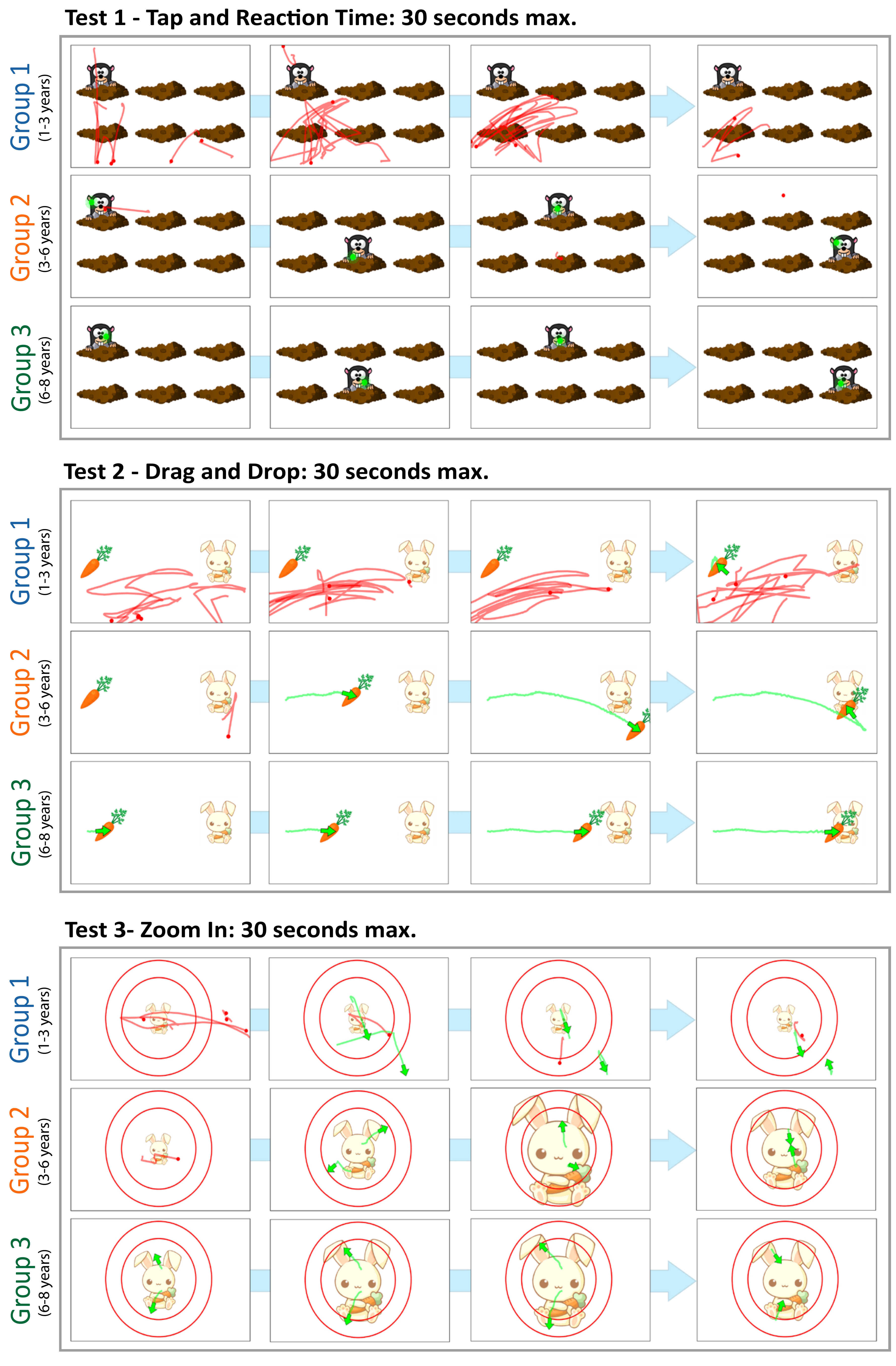}
\end{center}
\captionlistentry{}
\label{fig:figure_1}
\end{figure*}

\begin{figure*}[!]
\setcounter{figure}{0}
\begin{center}
   \includegraphics[width=\linewidth,height=\textheight-80pt]{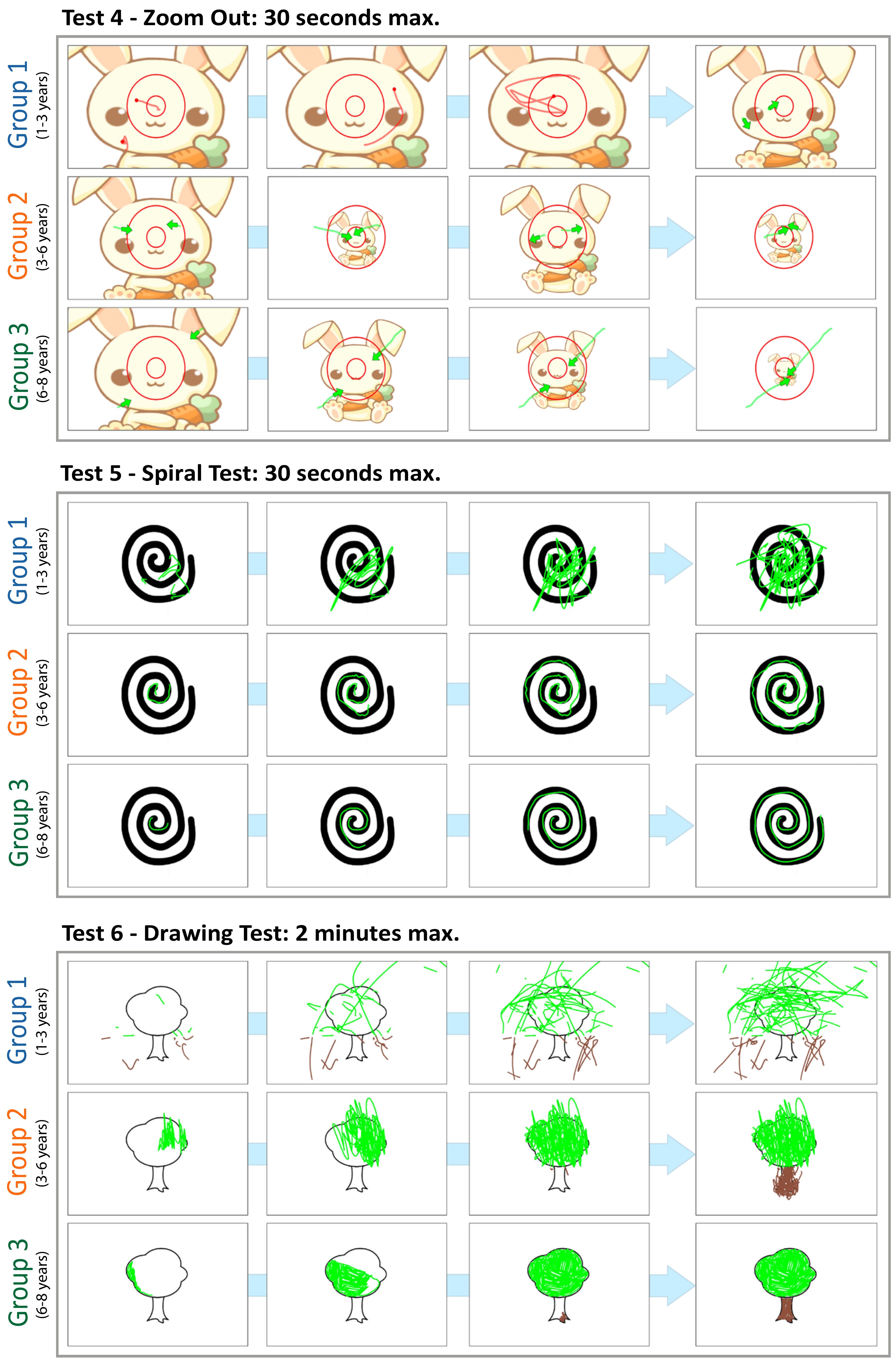}
\end{center}
    \caption{Examples of the ChildCIdb\_v1 tests performed by three different children age groups: \textbf{{\color[HTML]{1b62a5} Group 1}} (1 to 3 years), \textbf{{\color[HTML]{fd690f} Group 2}} (3 to 6 years), and \textbf{{\color[HTML]{1d6c11} Group 3}} (6 to 8 years). From Test 1-4, red marks indicate a poor interaction of the child compared to the expected in the test. Green marks indicate correct interaction. These marks are included here for a better comprehension. \textbf{Full video recordings of the different educational levels are available at \url{https://github.com/BiDAlab/ChildCIdb_v1}}
}
\end{figure*}

In particular, 6 different tests are considered in ChildCI, grouped in 2 main blocks: \textit{i)} touch, and \textit{ii)} stylus. Each one has a maximum amount of time to be performed and requires different levels of neuromotor and cognitive skills to be completed correctly. We briefly present next each of the tests:

\begin{itemize}
    \item \textit{Touch Block}
    \begin{itemize}
        \item \textbf{Test 1 - Tap and Reaction Time:} the screen shows 6 burrows and a single mole. When the children touch the mole using their finger it disappears from the current burrow and appears in another one at random. A total of 4 moles must be touched to finish the test. Just a single finger is needed to complete the test. This test requires fine motor skills (tap in a small area), as well as hand-eye coordination. The maximum time is 30 seconds.
        \item \textbf{Test 2 - Drag and Drop:} a carrot appears on the left side of the screen and a rabbit on the right. The aim is to touch the carrot, drag it from left to right and drop it on the rabbit. Just a single finger is needed to complete the test. This test combines fine motor skills (tap in a small area), pressure control, hand-eye coordination, and tracking of movement. The maximum time is 30 seconds.
        \item \textbf{Test 3 - Zoom In:} two red circles and a little rabbit appear on the screen. The children have to enlarge the rabbit and put it inside these circles for a short period of time. The rabbit can be only enlarged/shortened using two fingers. This test involves fine motor skills (put the rabbit inside two circles), coordination of the fingers (usually thumb and index finger) for the pinch movement, and accurate perception of the force. The maximum time is 30 seconds.
        \item \textbf{Test 4 - Zoom Out:} the goal is similar to Test 3. In this case, the rabbit is bigger and the children have to reduce its size to fit it inside the two red circles. Two fingers are needed to complete the test. This test requires the same motor and cognitive skills as Test 3 to be completed. The maximum time is 30 seconds.
    \end{itemize}
    \item \textit{Stylus Block}
    \begin{itemize}
        \item \textbf{Test 5 - Spiral Test:} a black spiral appears on the screen. The children, using the pen stylus, must draw along the spiral from the inner to the outer part, always trying to keep inside the black line that forms the spiral. This test requires precise hand-eye coordination, fine motor skills to control the stylus movement and follow a line without getting off the path, and visual tracking. The maximum time for this test is 30 seconds.
        \item \textbf{Test 6 - Drawing Test:} the outline of a tree appears on the screen and the children must colour it as well as they can. It involves hand-eye coordination, fine motor skills to control the stylus and stay within the outline of the tree, as well as planning and organization to colour it properly and fast. The maximum time is 2 minutes.
    \end{itemize}
\end{itemize}

Examples of the different tests can be seen in Fig.~\ref{fig:figure_1}, grouped by age. We include red and green marks along the tests to provide a better comprehension of the children interaction along the different age groups.

\section{Method}\label{sec4:method}

In order to shed some light on the questions considered in this study, i.e., \textit{i)} validate the discriminative power of the different tests included in ChildCI, and \textit{ii)} analyse whether there is any relationship between the chronological age of the children and their motor and cognitive development, the experimental framework of this study is carried out for the task of automatic children age group detection based on their motor and cognitive behaviors. Sec.~\ref{subsec41:features_sets} describes the feature set proposed for each of the tests considered in ChildCIdb\_v1. Sec.~\ref{subsec42:feature_selection} summarises the feature selection techniques considered. Finally, Sec.~\ref{subsec42:classification} indicates the different classification algorithms analysed.

\subsection{Feature Extraction}\label{subsec41:features_sets}

During the data collection process, each child performs the set of tests shown in Fig.~\ref{fig:figure_1}. This section presents the proposed feature sets for each test. In total, 111 global features are extracted referring to different types of skills.

\begin{itemize}
    \item \textbf{Test 1 - Tap and Reaction Time (Table~\ref{tab:tap-features}):} a set of 5 specific features is proposed.
    \item \textbf{Test 2 - Drag and Drop (Table~\ref{tab:drag-features}):} 28  features are proposed. In particular, 2 of them are proposed in this work and 26 are inspired on the study by~\cite{Hui2014}.
    \item \textbf{Test 3 and 4 - Zoom In/Out (Table~\ref{tab:zoom-features}):} a set of 20 features is proposed, 8 of them proposed in this work, 4 based on the study conducted in~\cite{Zaccagnino2021} and the remaining 8 inspired from~\cite{Hui2014}.
    \item \textbf{Test 5 - Spiral Test (Table~\ref{tab:spiral-features}):} for this test, 24 features are proposed. In particular, 3 of them are based on the study conducted in~\cite{Ishii2020}, 14 are inspired from~\cite{Sole-Casals2019}, 2 are based on the results obtained by~\cite{Lin2018} and the remaining 5 are proposed in this work.
    \item \textbf{Test 6 - Drawing Test:} we consider the 34 features originally presented in~\cite{Tolosana2021}.
\end{itemize}

The features studied in this work cover aspects such as: \textit{i)} the reaction time to touch the screen, \textit{ii)} the amount of time touching the target, \textit{iii)} the number of fingers used, and \textit{iv)} whether the child finishes the test (4 moles are touched, the carrot ends up on the rabbit, the rabbit is properly scaled, etc.), among others. In addition to the 111 global features presented above, 114 features based on preliminary studies in the field of HCI and related to Time, Kinematic, Direction, Geometry, and Pressure information are also considered~\cite{Tolosana2015, Marcos2008}, forming a final set of 225 global features in total.

\begin{table*}[!h]
    \parbox{.33\linewidth}{
        \centering
        \caption{Set of features proposed for ``Test 1: Tap and Reaction Time''.}
        \adjustbox{width=\textwidth/3, height=60pt}{
            \begin{tabular}{|c|c|}
                \hline
                \textbf{\#}                                      & \textit{\textbf{Feature Description}}                                                                 \\ \hline
                1 & \begin{tabular}[c]{@{}c@{}}Average Distance Between\rule{0pt}{10pt} \\ Tap and Centre of Mole\end{tabular}\\[+8pt] \hline
                2 & \begin{tabular}[c]{@{}c@{}}Standard Deviation Distance\rule{0pt}{10pt} \\ Between Tap and Centre of Mole\end{tabular} \\[+10pt] \hline
                3 & \begin{tabular}[c]{@{}c@{}}Maximum Distance Between Tap\rule{0pt}{10pt} \\ and Centre of Mole\end{tabular} \\[+8pt] \hline
                4 & \begin{tabular}[c]{@{}c@{}}Minimum Distance Between Tap\rule{0pt}{10pt} \\ and Centre of Mole\end{tabular} \\[+8pt] \hline
                5 & Number of Moles Touched\rule{0pt}{12pt} \\[+5pt] \hline
            \end{tabular}
        }
        \label{tab:tap-features}
    }
    \hfill
    \parbox{.65\linewidth}{
        \centering
        \caption{Set of features proposed for ``Test 2: Drag and Drop''. The features highlighted in \textbf{bold} are specifically designed for this test. The rest are inspired from~\cite{Hui2014}. LDP refers to the Largest Deviation Point.}
        \adjustbox{width=.655\textwidth, height=65pt}{
            \begin{tabular}{|c|c|c|c|}
                    \hline
                \# & \textbf{\textit{Feature Description}} & \# & \textbf{\textit{Feature Description}} \\ \hline
                1 & Average LDP Size & 2 & Standard Deviation LDP Size \\ \hline
                3 & Average LDP Velocity & 4 & Standard Deviation LDP Velocity \\ \hline
                5 & Average Start-to-LDP Latency (ms) & 6 & Standard Deviation Start-to-LDP Latency (ms) \\ \hline
                7 & Average Straight Start-to-LDP Length & 8 & Standard Deviation Straight Start-to-LDP Length \\ \hline
                9 & Average Start-to-LDP Direction & 10 & Standard Deviation Start-to-Stop Direction \\ \hline
                11 & Average Start-to-Stop Latency (ms) & 12 & Standard Deviation Start-to-Stop Latency (ms) \\ \hline
                13 & Average Straight Start-to-Stop Length & 14 & Standard Deviation Straight Start-to-Stop Length \\ \hline
                15 & Average Start-to-Stop Direction & 16 & Standard Deviation Start-to-Stop Direction \\ \hline
                17 & Average LDP-to-Stop Latency (ms) & 18 & Standard Deviation LDP-to-Stop Latency (ms) \\ \hline
                19 & Average Straight LDP-to-Stop Length & 20 & Standard Deviation LDP-to-Stop Length \\ \hline
                21 & Average LDP-to-Stop Direction & 22 & Standard Deviation LDP-to-Stop Direction \\ \hline
                23 & Average Start Point Velocity & 24 & Standard Deviation Start Point Velocity \\ \hline
                25 & Average Stop Point Velocity & 26 & Standard Deviation Stop Point Velocity \\ \hline
                27 & \begin{tabular}[c]{@{}c@{}}\textbf{The Carrot is Touched in the First} \\ \textbf{Pen-Down}\end{tabular} & 28 & \textbf{Carrot Ends Up in the Rabbit (Target)} \\ \hline
            \end{tabular}
        }
        \label{tab:drag-features}
    }
    \hfill
    \parbox{\linewidth}{
        \vspace{\baselineskip}
        \centering
        \caption{Set of features proposed for ``Test 3 and 4: Zoom-in and Zoom-out''. The features highlighted in \textbf{bold} are specifically designed for this test. The rest are inspired from~\cite{Zaccagnino2021} and~\cite{Hui2014}. Two fingers are needed for these tests. $FC$ refers to the first finger curve, $SC$ to the second finger curve, and $V$ to the velocity vector.}
        \adjustbox{width=0.75\textwidth}{
            \begin{tabular}{|c|c|c|c|}
                \hline
                \# & \textbf{\textit{Feature Description}} & \# & \textbf{\textit{Feature Description}} \\ \hline
                1 & \textbf{Total Time on Target (ms)} & 2 & \textbf{Reaction Time Until Using 2 Fingers (ms)} \\ \hline
                3 & \textbf{Maximum Scale} & 4 & \textbf{Minimum Scale} \\ \hline
                5 & \textbf{Average Scale} & 6 & \textbf{Standard Deviation Scale} \\ \hline
                7 & \textbf{\# Samples Using 2 Fingers} & 8 & \textbf{\# Samples Using 1 Finger} \\ \hline
                9 & Average $V_x$ $FC$ & 10 & Average $V_y$ $FC$ \\ \hline
                11 & Average $V_x$ $SC$ & 12 & Average $V_y$ $SC$ \\ \hline
                13 & $FC$ Trajectory Length & 14 & $SC$ Trajectory Length \\ \hline
                15 & $FC$ Trajectory Velocity & 16 & $SC$ Trajectory Velocity \\ \hline
                17 & Start Distance Between Both Fingers & 18 & Stop Distance Between Both Fingers \\ \hline
                19 & $FC$ Straight Length & 20 & $SC$ Straight Length \\ \hline
            \end{tabular}
        }
        \label{tab:zoom-features}
    }
    \hfill
    \parbox{\linewidth}{
        \vspace{\baselineskip}
        \centering
        \caption{Set of features proposed for ``Test 5: Spiral Test''. The features highlighted in \textbf{bold} are specifically designed for this test. The rest are inspired from~\cite{Ishii2020, Sole-Casals2019} and~\cite{Lin2018}. $R_{n}$ refers to the radial coordinates, $\theta_{n}$ to the angular coordinates, and $t_{n}$ to sample-to-sample times in seconds.}
        \adjustbox{width=0.80\textwidth}{
            \begin{tabular}{|c|c|c|c|}
                \hline
            \# & \textbf{Feature Description} & \# & \textbf{Feature Description} \\ \hline
            1 & Spiral Length & 2 & Average (Distance Between Points) \\ \hline
            3 & STD (Distance Between Points) & 4 & \textbf{Response Time (s)} \\ \hline
            5 & \begin{tabular}[c]{@{}c@{}}Sample Entropy (SENT)\\ {[}m=3, r=0.2{]}\end{tabular} & 6 & \begin{tabular}[c]{@{}c@{}}Mean Absolute Value (MAV)\\ $\frac{1}{N}\sum_{i=1}^{N}|R_{i}|$\end{tabular} \\ \hline
            7 & \begin{tabular}[c]{@{}c@{}}Variance (VAR)\\ $\frac{1}{N-1}\sum_{i=1}^{N}|R_{i}-\mu|^{2}$\end{tabular} & 8 & \begin{tabular}[c]{@{}c@{}}Root Mean Square (RMS)\\ $\sum_{i=1}^{N}\frac{1}{N}R_{i}^{2}$\end{tabular} \\ \hline
            9 & \begin{tabular}[c]{@{}c@{}}Log Detector (LOG)\\ $e^{\frac{1}{N}\sum_{i=1}^{N}log(|R_{i}|)}$\end{tabular} & 10 & \begin{tabular}[c]{@{}c@{}}Waveform Length (WL)\\ $\sum_{i=1}^{N-1}|R_{i+1}-R_{i}|$\end{tabular} \\ \hline
            11 & \begin{tabular}[c]{@{}c@{}}Standard Deviation (STD)\\ $\sqrt{\frac{1}{N-1}\sum_{i=1}^{N}|R_{i}-\mu|^{2}}$\end{tabular} & 12 & \begin{tabular}[c]{@{}c@{}}Difference Absolute Standard Deviation (ACC)\\ $\sqrt{\frac{1}{N-1}\sum_{i=1}^{N-1}|R_{i+1}-R_{i}|^{2}}$\end{tabular} \\ \hline
            13 & \begin{tabular}[c]{@{}c@{}}Fractal Dimension (FD)\\ {[}Higuchi's Algorithm with m=5{]}\end{tabular} & 14 & \begin{tabular}[c]{@{}c@{}}Maximum Fractal Length (MFL)\\ $log(\sum_{i=1}^{N-1}|R_{i+1}-R_{i}|)$\end{tabular} \\ \hline
            15 & \begin{tabular}[c]{@{}c@{}}Integrated EMG (IEMG)\\ $\sum_{i=1}^{N}|R_{i}|$\end{tabular} & 16 & \begin{tabular}[c]{@{}c@{}}Simple Square EMG (SSI)\\ $\sum_{i=1}^{N}R_{i}^{2}$\end{tabular} \\ \hline
            17 & \begin{tabular}[c]{@{}c@{}}Zero Crossing (ZC)\\ {[}\# Times the Signal Crosses Its Mean{]}\end{tabular} & 18 & \begin{tabular}[c]{@{}c@{}}Slope Sign Change (SSC)\\ {[}\# Times the Slope of the Sign Changes{]}\end{tabular} \\ \hline
            19 & \begin{tabular}[c]{@{}c@{}}Mean of Radial Difference Per Radian\\ $\frac{1}{N}\sum_{i=1}^{N-1}|\frac{R_{i+1}-R_{i}}{\theta_{i+1}-\theta_{i}}|$\end{tabular} & 20 & \begin{tabular}[c]{@{}c@{}}Mean of Radial Difference Per Second\\ $\frac{1}{N}\sum_{i=1}^{N-1}|\frac{R_{i+1}-R_{i}}{t_{i+1}-t_{i}}|$\end{tabular} \\ \hline
            21 & \textbf{\# Maximums in $R_{n}$} & 22 & \textbf{\# Minimums in $R_{n}$} \\ \hline
            23 & \textbf{Global Maximum Quartile of $R_{n}$} & 24 & \textbf{Global Minimum Quartile of $R_{n}$} \\ \hline
        \end{tabular}
    }
    \label{tab:spiral-features}
    }
\end{table*}

\subsection{Feature Selection}\label{subsec42:feature_selection}

The following feature selection techniques are used to choose the most discriminative features for each test from the total set originally extracted.

\begin{itemize}
    \item \textit{Sequential Forward Floating Search (SFFS):} is a widely used feature selection algorithm that searches for the best-correlated subset of features using specific optimization criteria. On the one hand, the solution is suboptimal because this algorithm does not take into account all possible combinations, but on the other hand, it does consider correlations between features, achieving high-accuracy results~\cite{Tolosana2015pre}. The implementation considered in this study has been provided by the MLxtend library\footnote{\url{http://rasbt.github.io/mlxtend/}}.
    
    \item \textit{Genetic Algorithm (GA):} is a metaheuristic algorithm based on Charles Darwin's theory of evolution. It is presented in our previous work~\cite{Tolosana2021} and is mainly inspired by the natural selection process of evolution, where over generations and by using operators such as mutation, crossover, and selection, a positive evolution towards better solutions occurs. It is widely used as a feature selection method as it reduces computational time, improves prediction performance, and allows for a better understanding of data~\cite{Chandrashekar2014, Saibene2023}. Our public version of this library can be found on GitHub\gafootnote. In our experiments, we have considered the parameters that provided the best performance during the development stage: an initial population = 200, a random number of generations = 100, a crossover rate = 0.6, and a mutation rate = 0.05.
\end{itemize}

\subsection{Classification Algorithms}\label{subsec42:classification}

All classifiers are publicly available on Scikit-Learn\footnote{\url{https://scikit-learn.org/stable/}}. The parameters used for each classifier are those with the best performance during the development stage.

\begin{itemize}
    \item \textit{Support Vector Machines (SVM):} this algorithm builds a hyperplane or set of hyperplanes in a high- or infinite-dimensional space that differentiates the classes as well as possible. In our case, the regularization parameter is 0.1, the kernel type is ``polynomial'' with 3 degrees and the coefficient is ``scaled''.
    
    \item \textit{Random Forest (RF):} this is an ensemble method consisting of a defined number of small decision trees, called estimators. A combination of the estimator's decisions is produced to get a more accurate prediction. In our experiments, the number of estimators is 10, the function to measure the quality of a split is ``gini'' and the maximum depth of the tree is 75.
\end{itemize}

The SVM and RF classifiers are selected in this study due to their popularity in several machine learning tasks. They offer high versatility and solid results, as can be seen in~\cite{Tolosana2021}, outperforming other machine learning approaches.

\begin{table*}[t]
    \caption{Results achieved in terms of Accuracy (\%) over the final evaluation dataset of ChildCIdb\_v1 for each of the tests considered. Standard deviations are reported in brackets. We highlight in \textbf{bold} the feature selector and classifier with the best results.}
    \label{tab:accuracy}
    \resizebox{\linewidth}{!}{%
        \begin{tabular}{ll|ll|ll}
        \multicolumn{1}{l}{} & \multicolumn{1}{c|}{} & \multicolumn{2}{c|}{\textbf{SFFS}} & \multicolumn{2}{c}{\textbf{GA}} \\ \cline{3-6} 
        \multicolumn{1}{l}{} &  & \multicolumn{1}{c}{\textbf{SVM}}\rule{0pt}{12pt} & \multicolumn{1}{c|}{\textbf{RF}} & \multicolumn{1}{c}{\textbf{SVM}} & \multicolumn{1}{c}{\textbf{RF}} \\ \hline
        \multicolumn{1}{c|}{\multirow{4}{*}{\textit{Finger}}} & \begin{tabular}[c]{@{}l@{}}Test 1: Tap and Reaction\rule{0pt}{15pt}\end{tabular} & \textbf{89.79 (\boldmath$\pm$2.19e-02)} & 87.71 ($\pm$2.41e-02) & 88.51 ($\pm$2.52e-02) & 85.00 ($\pm$2.25e-02) \\
        \multicolumn{1}{c|}{} & Test 2: Drag and Drop & \textbf{87.23 (\boldmath$\pm$2.87e-02)} & 83.89 ($\pm$4.77e-02) & 85.47 ($\pm$2.29e-02) & 83.09 ($\pm$3.95e-02) \\
        \multicolumn{1}{c|}{} & Test 3: Zoom In & 80.22 ($\pm$2.55e-02) & 78.14 ($\pm$2.05e-02) & \textbf{81.33 (\boldmath$\pm$9.94e-03)} & 78.15 ($\pm$2.49e-02) \\
        \multicolumn{1}{c|}{} & Test 4: Zoom Out & 79.74 ($\pm$2.08e-02) & 79.26 ($\pm$5.32e-02) & \textbf{82.45 (\boldmath$\pm$6.06e-02)} & 80.22 ($\pm$4.05e-02) \\[+4pt] \cline{1-6}
        \multicolumn{1}{c|}{\multirow{2}{*}{\textit{Stylus}}} & Test 5: Spiral Test\rule{0pt}{15pt} & \textbf{88.67 (\boldmath$\pm$2.34e-02)} & 82.93 ($\pm$1.95e-02) & 85.16 ($\pm$1.90e-02) & 81.33 ($\pm$2.10e-02) \\
        \multicolumn{1}{c|}{} & Test 6: Drawing Test & \textbf{90.45 (\boldmath$\pm$4.47e-02)} & 88.69 ($\pm$5.09e-02) & 81.51 ($\pm$5.56e-02) & 80.37 ($\pm$5.04e-02) \\[+5pt] \hline
        \end{tabular}
    }
\end{table*}

\begin{figure*}[!t]
    \centering
    \parbox{0.40\linewidth}{
        \vspace{0.15cm}
        \centering
        \includegraphics[width=0.40\textwidth]{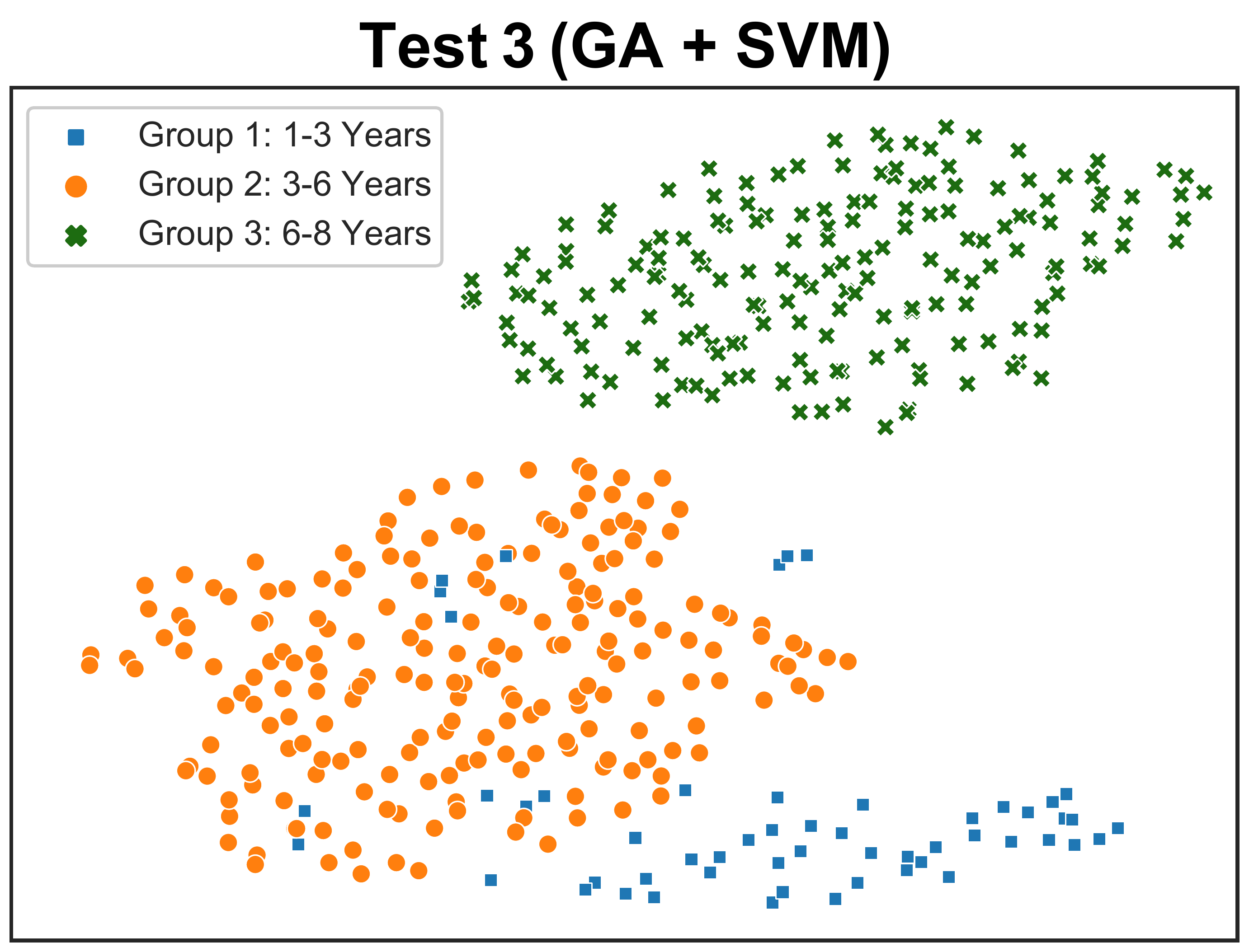}
        \label{fig:zoom-in-umap}
    }
    \parbox{0.40\linewidth}{
        \vspace{0.15cm}
        \centering
        \includegraphics[width=0.40\textwidth]{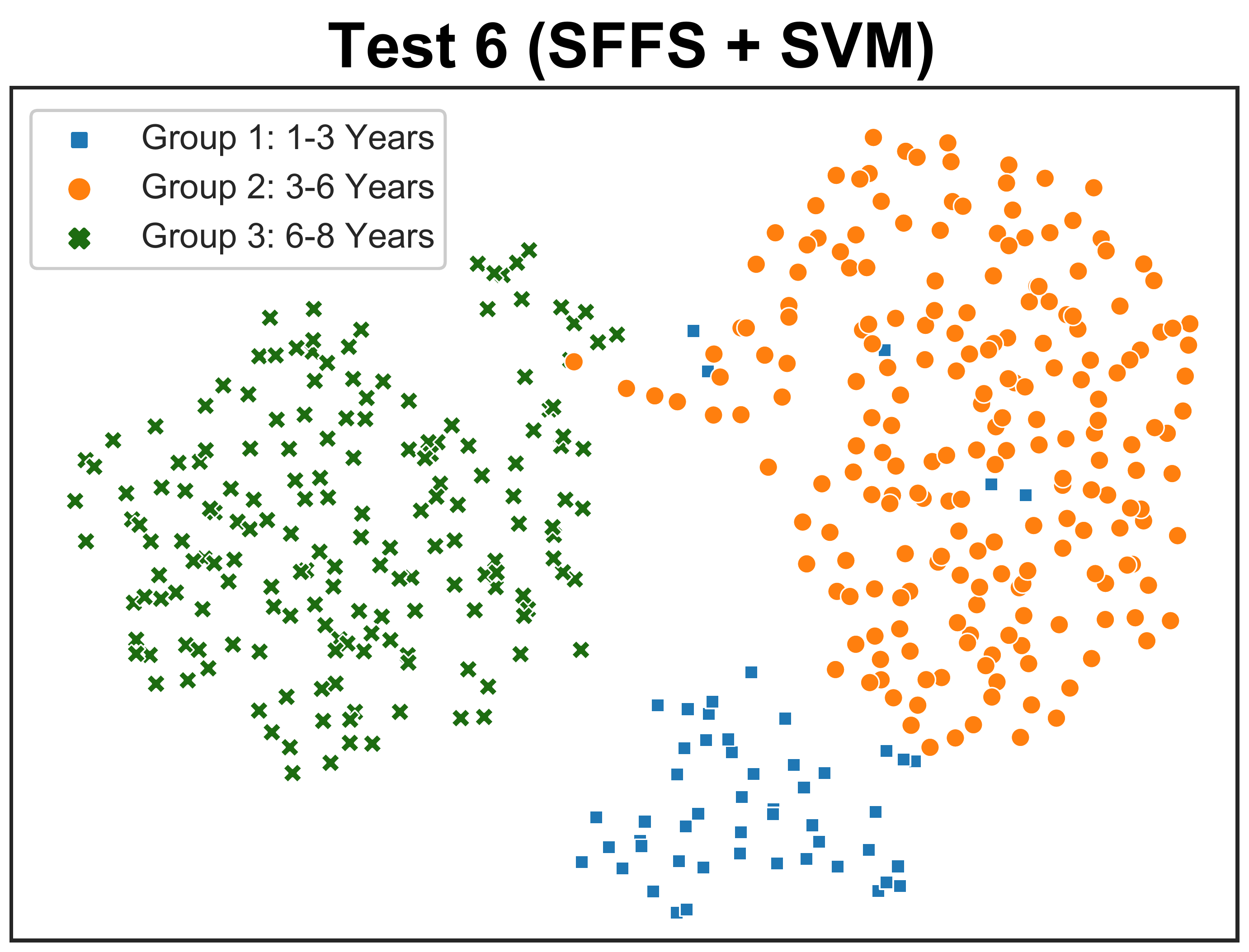}
        \label{fig:tree-umap}
    }
    \caption{Examples of children age groups formed based on the motor and cognitive features proposed in this study. Three groups can be observed in each graph: \textbf{{\color[HTML]{1b62a5} Group 1}} (1 to 3 years), \textbf{{\color[HTML]{fd690f} Group 2}} (3 to 6 years), and \textbf{{\color[HTML]{1d6c11} Group 3}} (6 to 8 years). Each point refers to one child of ChildCIdb\_v1.}
    \label{fig:umap}
\end{figure*}

\section{Experiments and results}\label{sec5:experiments}

\subsection{Experimental Protocol}\label{subsec51:protocol}

The experimental protocol considered in this work is designed with the aim of age group detection based on the children interaction behavior. The following 3 different age groups are considered: \textbf{{\color[HTML]{1b62a5} Group 1}} (children aged 1 to 3 years), \textbf{{\color[HTML]{fd690f} Group 2}} (children aged 3 to 6 years), and  \textbf{{\color[HTML]{1d6c11} Group 3}} (children aged 6 to 8 years). It is important to remark that this age categorization differs from Piaget's stages as we focus on specific motor-cognitive skills (tap, drag and drop, pinch, etc.) rather than on more generic skills presented by Piaget's levels. This decision is also supported by previous approaches in the literature that correlate children's gestures with ages~\cite{Crescenzi2019, vatavu2015touch}, and by the neurologists, psychologist, and educators of GSD School during the acquisition of the database. ChildCIdb\_v1 is divided into 2 data subsets: development (80\%) and evaluation (20\%). The development dataset is used for the training of the age group detection systems whereas the evaluation dataset is used to test the performance of the trained systems, excluding the children considered in the development dataset. In addition, and only during the development stage, a data augmentation technique is used as the data available in Groups 1 and 3 are smaller than in Group 2. This technique is called SMOTE and is publicly available in the Imbalanced-Learn toolbox\footnote{\url{https://imbalanced-learn.org/stable/}}. To provide a better analysis of the results, $k$-fold cross-validation with $k$=5 is used, showing the final evaluation results of the 5-fold cross-validation. All experiments are run on a machine with an Intel i7-9700 processor and 32GB of RAM.

\begin{figure*}[!t]
    \begin{center}
       \includegraphics[width=0.8\linewidth]{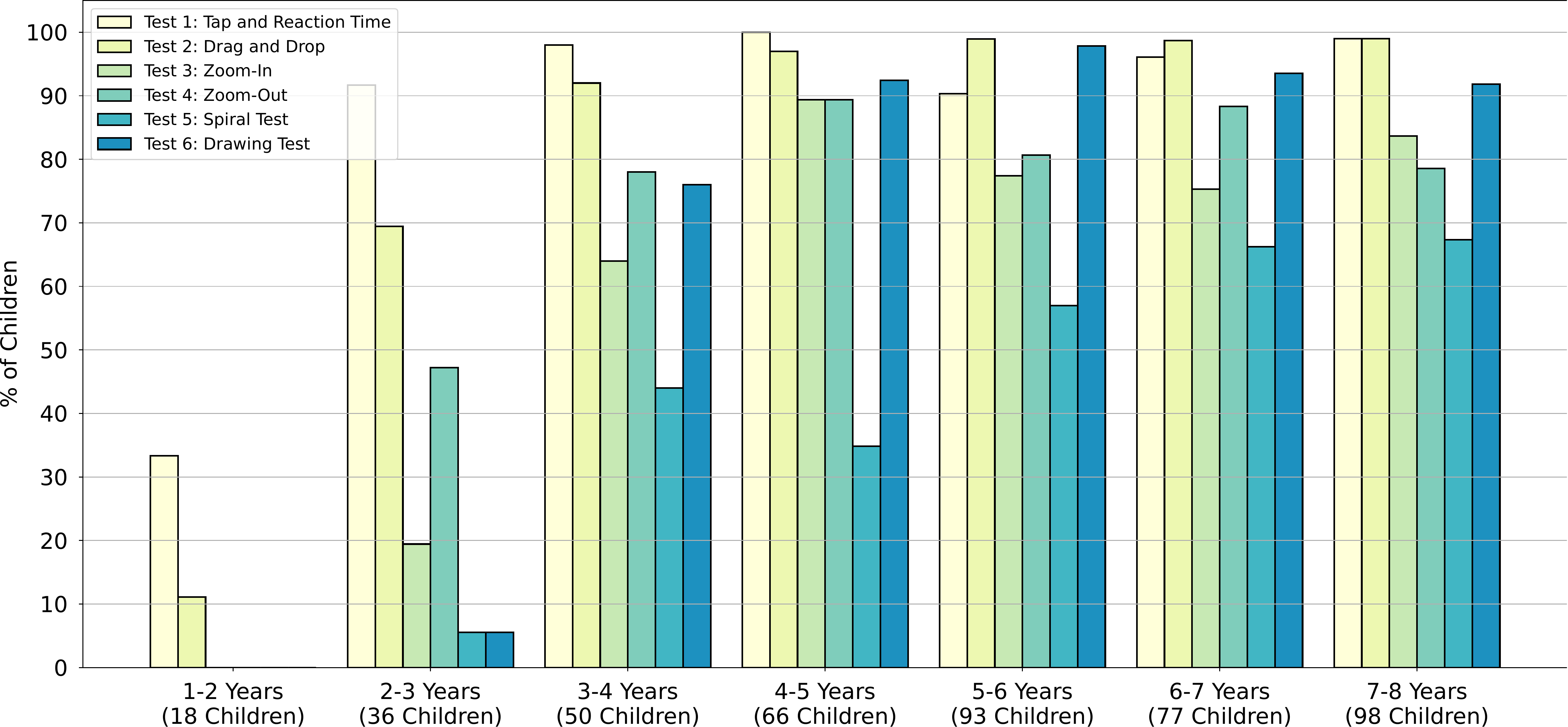}
    \end{center}
    \caption{Percentage of ChildCI tests completed for the 438 children data captured in ChildCIdb\_v1 by their chronological ages.}
    \label{fig:types_of_gestures}
\end{figure*}

\subsection{Experimental Results}\label{subsec52:results}

This section analyses the performance of the methods presented above to the children age group detection task based on motor and cognitive behaviors when interacting with mobile devices. The analysis is carried out in two different stages: \textit{i)} a first test-by-test analysis is performed, and \textit{ii)} then a combination of tests is conducted to analyse the potential of the ChildCI tests as a whole. The results obtained are measured in terms of Accuracy (\%).

\begin{enumerate}

    \item \textbf{Single-Test Scenario:} Table~\ref{tab:accuracy} shows the results obtained in each test using the different classifiers and feature selectors considered. 
    
    We first analyse the results test by test using the best results achieved for each one. As can be seen, Test 6 obtains the best accuracy (90.45\%), while the worst result is for Test 3 (81.33\%). Children can draw the tree in many different ways. Therefore, in Test 6 they have more freedom to interact with the device and, as a result, the variation between groups can be better observed, leading to high accuracy in the age group detection task. Analysing the results according to the classifier studied, SVM always achieves better results than RF, regardless of the feature selector used. In particular, SVM achieves an average accuracy of over 86\%, while for RF the average is less than 83\%. It is also interesting to analyse the results by the type of feature selectors considered. In most cases, SFFS provides the best results, achieving in Test 1, Test 2, Test 5 and Test 6 rates above 87\% accuracy. Nevertheless, GA performs better for Tests 3 and 4, reaching 81.33\% and 82.45\% accuracy. This proves the potential of our proposed feature selector algorithm that is publicly available\gafootnote. In addition, we analyse the results according to the writing input used (stylus/finger). Always looking at the best tests for each input method, the results obtained in terms of accuracy are similar, indicating that the input method used is not really relevant for the age group detection task.
    
    In order to show the results in a more visual way, we use a popular visualization technique, called Uniform Manifold Approximation and Projection (UMAP)~\cite{mcinnes2018umap}. Fig.~\ref{fig:umap} shows 2 examples of the children age groups formed based on the motor and cognitive features proposed in this study. Therefore, those children belonging to the same age group and whose motor and cognitive interactions on each test are similar should appear in the same group and placed contiguously to each other. As can be seen in both cases, there is a point cloud with 3 distinct groups, indicating a high correlation between the age of the children and the way they interact with mobile devices. However, some children are in a different point cloud to their own age group. For example, in the results of Test 6 in Fig.~\ref{fig:umap}, we can see that there are children from \textbf{{\color[HTML]{1b62a5} Group 1}} who are in the point cloud of \textbf{{\color[HTML]{fd690f} Group 2}}. These particular cases could be an indicator that these children have more advanced motor and cognitive aspects than their age group.

    In view of the results obtained, we can shed some light on the key questions analysed in this study. First, the validation of the discriminative power of the different tests included in the ChildCI framework. The results achieved in Table~\ref{tab:accuracy} prove that ChildCI tests are able to measure different children motor and cognitive features for the different ages. Second, the analysis of whether there is any relationship between the chronological age of the children and their motor and cognitive development. The point clouds shown in Fig.~\ref{fig:umap} indicate that there seems to be a good relationship between the motor and cognitive features proposed in this study and the chronological age of the children.

    \begin{table*}[t]
        \caption{Best average results in terms of Accuracy (\%) for all the possible combinations, in groups of 2, 3, 4, 5, and 6 tests ($C_{6,2}$, $C_{6,3}$, $C_{6,4}$, $C_{6,5}$ and $C_{6,6}$), of the different ChildCIdb\_v1 tests. Standard deviations are reported in brackets. We highlight in \textbf{bold} the combination with the best result.}
        \label{tab:combinations}
        \adjustbox{width=1\textwidth}{
            \begin{tabular}{c|c|c|c|c|c|c|c}
            \textbf{Combination} & \textbf{\begin{tabular}[c]{@{}c@{}}Test 1: \\ Tap and Reaction Time\end{tabular}} & \textbf{\begin{tabular}[c]{@{}c@{}}Test 2:\\Drag and Drop\end{tabular}} & \textbf{\begin{tabular}[c]{@{}c@{}}Test 3:\\Zoom In\end{tabular}} & \textbf{\begin{tabular}[c]{@{}c@{}}Test 4:\\Zoom Out\end{tabular}} & \textbf{\begin{tabular}[c]{@{}c@{}}Test 5:\\Spiral Test\end{tabular}} & \textbf{\begin{tabular}[c]{@{}c@{}}Test 6:\\Drawing Test\end{tabular}} & \textbf{Accuracy (\%)} \\ \hline
             C2 &  &  &  & \cmark &  & \cmark & 89.59 $(\pm 2.64)$ \\
             C3 &  &  &  & \cmark & \cmark & \cmark & 91.40 $(\pm 2.47)$ \\
             C4 &  &  & \cmark & \cmark & \cmark & \cmark & 91.68 $(\pm 2.75)$ \\
             C5 &  \cmark &  & \cmark & \cmark & \cmark & \cmark & 92.86 $(\pm 2.19)$ \\
             \textbf{C6} &  \cmarkbig & \cmarkbig & \cmarkbig & \cmarkbig & \cmarkbig & \cmarkbig & \textbf{93.08 \boldmath$(\pm 2.37)$} \\ \hline
            \end{tabular}
        }
    \end{table*}

    \begin{table}[t]
        \caption{Pairwise comparison results with Mann-Whitney U test and Bonferroni correction for the combinations of tests.}
        \label{tab:mann-whitney}
        \centering
        \adjustbox{width=0.49\textwidth}{
            \begin{tabular}{c|c|c}
            \textbf{\begin{tabular}[c]{@{}c@{}}Pairwise \\ Comparison\end{tabular}} & \textbf{\begin{tabular}[c]{@{}c@{}}p-value \\ (Mann-Whitney U test)\end{tabular}} & \textbf{\begin{tabular}[c]{@{}c@{}}\boldmath$H_0$ is rejected\\ (\boldmath$< \alpha_{cor} = 0.01$)\end{tabular}}\\ \hline
            C2 vs C3 & \multicolumn{1}{c|}{0.17} & No \\
            C2 vs C4 & \multicolumn{1}{c|}{0.42} & No \\
            C2 vs C5 & \multicolumn{1}{c|}{1.71e-04} & Yes \\
            C2 vs C6 & \multicolumn{1}{c|}{6.97e-05} & Yes \\ \hline
            C3 vs C4 & \multicolumn{1}{c|}{0.58} & No \\
            C3 vs C5 & \multicolumn{1}{c|}{6.27e-04} & Yes \\
            C3 vs C6 & \multicolumn{1}{c|}{1.89e-03} & Yes \\ \hline
            C4 vs C5 & \multicolumn{1}{c|}{1.32e-03} & Yes \\ 
            C4 vs C6 & \multicolumn{1}{c|}{5.19e-04} & Yes \\ \hline
            C5 vs C6 & \multicolumn{1}{c|}{0.51} & No \\ \hline
            \end{tabular}
        }
    \end{table}
    
    Finally, for completeness, we analyse in Fig.~\ref{fig:types_of_gestures} the type of gestures and tests children are able to perform according to their chronological age. Test 6 (Drawing Test) is considered correctly completed when at least 70\% of the tree surface is coloured. Looking at those gestures performed with the finger (from Test 1 to 4), we can see how gestures such as tap or drag and drop are easily achievable from the age of 2-3 years. However, more complex gestures such as zoom-in (Test 3) and zoom-out (Test 4) are strongly influenced by the age of the child, as until the age of 3-5 years old they are not able to complete them in general. Similar conclusions can be observed for those cases where the child interacts with the stylus (Test 5 and Test 6), mainly due to the fine motor skills needed to perform this type of tests.
    
    \item \textbf{Multiple-Test Scenario:} In ChildCI framework there are 6 tests in total, with the corresponding best machine learning models for each test (previous experiment). The present experiment analyses the potential of combining these tests. We consider combinatorial operations with all possible test combinations. The number of combinations is indicated by the following equation:
    
    \begin{equation}
    \centering
    C_{n,x} = \binom{n}{x} = \frac{n!}{x!(n-x)!}
    \end{equation}
    
    The number of total observations is represented by $n$ whereas $x$ refers to the number of selected elements. We combine the individual tests into groups of 2, 3, 4, 5, and 6 tests (all tests together). In total there are 57 possible combinations: \textit{i)} 15 combinations in groups of 2 tests, \textit{ii)} 20 combinations in groups of 3 tests, \textit{iii)} 15 combinations in groups of 4 tests, \textit{iv)} 6 combinations in groups of 5 tests, and \textit{v)} 1 combination of all tests together. In particular, the machine learning models trained for each test during the development stage generate 3 probabilities (one for each children age group, values between 0 and 1) whose sum cannot exceed 1. To combine the different tests, a majority voting ensemble is considered. Then, the associated age group of a child is determined by the highest number of votes among the classifiers. For tie-breakers, an average of the probabilities generated for each set of grouped tests is calculated, with the highest value determining the age group associated with the child.

    In order to generate reliable results, 25 random \textit{k}-fold cross-validation are performed ($k=5$). Table~\ref{tab:combinations} shows the best average results in terms of accuracy for each group of test combinations (from 2-test to 6-test). As we can observe, the more tests are considered, the better the results are. In particular, the combination of all tests (C6) offers the best result, reaching 93\% accuracy. Therefore, this result proves that i) all tests considered in ChildCI frameworks are valuable, and ii) a combination of tests is a good practice to obtain better results for the children age group detection task. Finally, for completeness, we include a statistical analysis through the Kruskal-Wallis test among the 5 combinations presented in Table~\ref{tab:combinations} to check whether the results are statistically significant. We propose the following hypothesis: \\

    \begin{itemize}
        \item $H_0$ (Null Hypothesis): There are no significant differences between the test combinations.
        \item $H_1$ (Alternative Hypothesis): There are significant differences between the test combinations.
    \end{itemize}
    
    The Kruskal-Wallis test shows a p-value of 6.24e-06. Therefore, setting a significance value $\alpha=0.05$ and a Bonferroni-corrected significance value ($\alpha_{cor} = 0.05/5 = 0.01$), $H_0$ can be rejected (p-value $< \alpha_{cor}$), and a significant difference between the mean accuracies of the combinations is demonstrated. Based on these results, we applied a post hoc Mann-Whitney U test to find out between which pairs of combinations there are significant differences in the results. Table~\ref{tab:mann-whitney} shows the results of the Mann-Whitney U test and if the null hypothesis $H_0$ is rejected. As can be seen, in all pairwise comparisons related to the C6 combination (combination of all tests) there are significant differences in the results obtained, with the exception of the comparison with C5, where it appears that including Test 2 (Drag and Drop) is not significant for an $\alpha_{cor}$ value of 0.01.

\end{enumerate}

\section{Conclusion and Future Work}\label{sec6:conclusion_future_work}

The proposal of automatic methods that quantify the motor and cognitive development of the children through the interaction with mobile devices is one of the main motivations of our ChildCI framework. As a first step to reach that future goal, this study proposes a comprehensive analysis evaluating the discriminative power of the tests presented in our ChildCI framework, and an analysis of whether there is any relationship between the chronological age of the children and their motor and cognitive development. 

For each of the tests considered, a robust set of features representing cognitive and motor aspects of children during interaction with mobile devices is presented. The experimental framework of this study is carried out for the automatic children age group detection task based on similar motor and cognitive behaviors.

The results achieved shed some light on the questions and contributions analysed in this study. Indeed, there is a relationship between children's chronological age, their motor and cognitive development and the type of test they are able to perform when interacting with mobile devices. Fig.~\ref{fig:umap} shows a high correlation between the age of the children and the way they interact with the devices, denoting the way in which the children perform the tests can give a rough indication of their chronological age group. Nevertheless, 100\% accuracy is not achieved in the age group detection task because children's evolution is a maturation process. This means that children of the same age group may have more/less advanced motor and cognitive aspects depending on their development, as can be seen in Fig.~\ref{fig:types_of_gestures}.

In addition, the potential and discriminative power of the tests included in the ChildCI framework is proved. The results achieved in Table~\ref{tab:accuracy} and Table~\ref{tab:combinations} demonstrate that ChildCI tests are able to measure different children motor and cognitive features for the different ages. This indicates both the correct design of the tests, discussed and approved by specialists such as neurologists, child psychologists and educators, and their inherent applicability to other research problems around e-Learning and e-Health. In particular, such kind of applications could include: \textit{i)} serving as a precise and personalized assessment tool to detect delays or difficulties in children's motor and cognitive development, enabling early interventions~\cite{Acien2022, Gomez2023}; \textit{ii)} providing a platform for the development of personalized e-learning applications~\cite{Daza2022}, by adapting the content and challenges to individual needs of children, among others.

Future works will be oriented towards: \textit{i)} relating children's interaction information with mobile devices to the other metadata stored in ChildCIdb (school grades, ADHD, previous experience using mobile devices, prematurity, etc.), \textit{ii)} presentation of new versions of the database analysing longitudinally the evolution of children when performing the different ChildCIdb tests, and \textit{iii)} take advantage of ChildCIdb's potential in other e-Health and e-Learning research areas and problems.

\section*{Acknowledgements}

This work has been supported by projects: INTER-ACTION (PID2021-126521OB-I00 MICINN/FEDER) and HumanCAIC (TED2021-131787B-I00 MICINN). J.C. Ruiz-Garcia is supported by the Madrid Government (Comunidad de Madrid-Spain) under the Multiannual Agreement with Autonomous University of Madrid in the line Encouragement of the Research of Young Researchers, in the context of the V PRICIT (Regional Programme of Research and Technological Innovation). This is an on-going project carried out with the collaboration of the school GSD Las Suertes in Madrid, Spain.

{
\bibliographystyle{IEEEtran}
\bibliography{ChildCI_2}

\begin{thebibliography}{10}
\providecommand{\url}[1]{#1}
\csname url@samestyle\endcsname
\providecommand{\newblock}{\relax}
\providecommand{\bibinfo}[2]{#2}
\providecommand{\BIBentrySTDinterwordspacing}{\spaceskip=0pt\relax}
\providecommand{\BIBentryALTinterwordstretchfactor}{4}
\providecommand{\BIBentryALTinterwordspacing}{\spaceskip=\fontdimen2\font plus
\BIBentryALTinterwordstretchfactor\fontdimen3\font minus
  \fontdimen4\font\relax}
\providecommand{\BIBforeignlanguage}[2]{{%
\expandafter\ifx\csname l@#1\endcsname\relax
\typeout{** WARNING: IEEEtran.bst: No hyphenation pattern has been}%
\typeout{** loaded for the language `#1'. Using the pattern for}%
\typeout{** the default language instead.}%
\else
\language=\csname l@#1\endcsname
\fi
#2}}
\providecommand{\BIBdecl}{\relax}
\BIBdecl

\bibitem{Antle2021}
A.~N. Antle and J.~P. Hourcade, ``{Research in Child–Computer Interaction:
  Provocations and Envisioning Future Directions},'' \emph{International
  Journal of Child-Computer Interaction}, p. 100374, 2021.

\bibitem{Kabali2015}
H.~K. Kabali, M.~M. Irigoyen, R.~Nunez-Davis, J.~G. Budacki, S.~H. Mohanty,
  K.~P. Leister, and J.~Bonner, Robert~L., ``{Exposure and Use of Mobile Media
  Devices by Young Children},'' \emph{Pediatrics}, vol. 136, no.~6, pp.
  1044--1050, 2015.

\bibitem{kilicc2019exposure}
A.~O. K{\i}l{\i}{\c{c}}, E.~Sari, H.~Yucel, M.~M. O{\u{g}}uz, E.~Polat, E.~A.
  Acoglu, and S.~Senel, ``{Exposure to and Use of Mobile Devices in Children
  Aged 1-60 Months},'' \emph{European Journal of Pediatrics}, vol. 178, no.~2,
  pp. 221--227, 2019.

\bibitem{antle2020child}
A.~N. Antle and C.~Frauenberger, ``{Child-Computer Interaction in Times of a
  Pandemic},'' \emph{International Journal of Child-Computer Interaction}, p.
  100201, 2020.

\bibitem{Tolosana2021}
R.~Tolosana, J.~C. Ruiz-Garcia, R.~Vera-Rodriguez, J.~Herreros-Rodriguez,
  S.~Romero-Tapiador, A.~Morales, and J.~Fierrez, ``{Child-Computer Interaction
  with Mobile Devices: Recent Works, New Dataset, and Age Detection},''
  \emph{IEEE Transactions on Emerging Topics in Computing}, pp. 1--1, 2022a.

\bibitem{piaget2008}
J.~Piaget and B.~Inhelder, \emph{{The Psychology of the Child}}.\hskip 1em plus
  0.5em minus 0.4em\relax Basic books, 2008.

\bibitem{Morante2016}
M.~Morante, M.~Costa, and N.~Rodriguez, ``{Children's Evolving Capabilities in
  Their Interaction with Touchable Devices from Birth to 2 Years Old},'' in
  \emph{Proc. 15th International Conference on Interaction Design and
  Children}, 2016.

\bibitem{Hourcade2015}
J.~P. Hourcade, S.~L. Mascher, D.~Wu, and L.~Pantoja, ``{Look, My Baby Is Using
  an IPad! An Analysis of YouTube Videos of Infants and Toddlers Using
  Tablets},'' in \emph{Proc. 33rd Annual ACM Conference on Human Factors in
  Computing Systems}, 2015.

\bibitem{vatavu2015touch}
R.-D. Vatavu, G.~Cramariuc, and D.~M. Schipor, ``{Touch Interaction for
  Children Aged 3 to 6 Years: Experimental Findings and Relationship to Motor
  Skills},'' \emph{International Journal of Human-Computer Studies}, vol.~74,
  pp. 54--76, 2015.

\bibitem{Vera-Rodriguez2020}
R.~Vera-Rodriguez, R.~Tolosana, J.~Hernandez-Ortega, A.~Acien, A.~Morales,
  J.~Fierrez, and J.~Ortega-Garcia, ``{Modeling the Complexity of Signature and
  Touch-Screen Biometrics using the Lognormality Principle},'' in \emph{The
  Lognormality Principle and its Applications in e-Security, e-Learning and
  e-Health}, R.~Plamondon, A.~Marcelli, and M.~Ángel Ferrer, Eds., 2020, pp.
  65--86.

\bibitem{nacher2015multi}
V.~Nacher, J.~Jaen, E.~Navarro, A.~Catala, and P.~Gonz{\'a}lez, ``{Multi-Touch
  Gestures for Pre-Kindergarten Children},'' \emph{International Journal of
  Human-Computer Studies}, vol.~73, pp. 37--51, 2015.

\bibitem{chen2020examining}
Z.~Chen, Y.-P. Chen, A.~Shaw, A.~Aloba, P.~Antonenko, J.~Ruiz, and L.~Anthony,
  ``{Examining the Link between Children's Cognitive Development and
  Touchscreen Interaction Patterns},'' in \emph{Proc. International Conference
  on Multimodal Interaction}, 2020.

\bibitem{Tolosana2021DeepWrite}
R.~Tolosana, P.~Delgado{-}Santos, A.~Perez{-}Uribe, R.~Vera{-}Rodriguez,
  J.~Fierrez, and A.~Morales, ``{DeepWriteSYN: On-Line Handwriting Synthesis
  via Deep Short-Term Representations},'' in \emph{Proc. 35th AAAI Conference
  on Artificial Intelligence}, 2021.

\bibitem{Tolosana2022SVC}
R.~Tolosana and et~al., ``{SVC-onGoing: Signature Verification Competition},''
  \emph{Pattern Recognition}, 2022b.

\bibitem{Price2015}
S.~Price, C.~Jewitt, and L.~Crescenzi, ``{The Role of iPads in Pre-School
  Children's Mark Making Development},'' \emph{Computers \& Education},
  vol.~87, pp. 131--141, 2015.

\bibitem{remi2015exploring}
\BIBentryALTinterwordspacing
C.~R{\'e}mi, J.~Vaillant, R.~Plamondon, L.~Prevost, and T.~Duval, ``{Exploring
  the Kinematic Dimensions of Kindergarten Children's Scribbles},'' in
  \emph{Proc. Conference of the International Graphonomics Society}, 2015.
  [Online]. Available: \url{https://hal.univ-antilles.fr/hal-01165910}
\BIBentrySTDinterwordspacing

\bibitem{tabatabaey2015analyses}
N.~Tabatabaey-Mashadi, R.~Sudirman, R.~M. Guest, and P.~I. Khalid, ``{Analyses
  of Pupils' Polygonal Shape Drawing Strategy with Respect to Handwriting
  Performance},'' \emph{Pattern Analysis and Applications}, vol.~18, no.~3, pp.
  571--586, 2015.

\bibitem{Singh2023}
R.~Singh, A.~{Kumar Singh Kushwaha}, Chandni, and R.~Srivastava, ``{Recent
  Trends in Human Activity Recognition – A Comparative Study},''
  \emph{Cognitive Systems Research}, vol.~77, pp. 30--44, 2023.

\bibitem{Schadenberg2022}
B.~Schadenberg, M.~Neerincx, F.~Cnossen, and R.~Looije, ``{Personalising Game
  Difficulty to Keep Children Motivated to Play with a Social Robot: A Bayesian
  Approach},'' \emph{Cognitive Systems Research}, vol.~43, pp. 222--231, 2017.

\bibitem{Ishii2020}
N.~Ishii, Y.~Mochizuki, K.~Shiomi, M.~Nakazato, and H.~Mochizuki, ``{Spiral
  drawing: Quantitative analysis and artificial-intelligence-based diagnosis
  using a smartphone},'' \emph{Journal of the Neurological Sciences}, vol. 411,
  p. 116723, 2020.

\bibitem{Sole-Casals2019}
J.~Solé-Casals, I.~Anchustegui-Echearte, P.~Marti-Puig, P.~M. Calvo,
  A.~Bergareche, J.~I. Sánchez-Méndez, and K.~Lopez-de Ipina, ``{Discrete
  Cosine Transform for the Analysis of Essential Tremor},'' \emph{Frontiers in
  Physiology}, vol.~9, 2019.

\bibitem{Lin2018}
P.-C. Lin, K.-H. Chen, B.-S. Yang, and Y.-J. Chen, ``{A Digital Assessment
  System for Evaluating Kinetic Tremor in Essential Tremor and Parkinson's
  Disease},'' \emph{BMC Neurology}, vol.~18, no.~1, p.~25, 2018.

\bibitem{Hui2014}
H.~Xu, Y.~Zhou, and M.~R. Lyu, ``{Towards Continuous and Passive Authentication
  via Touch Biometrics: An Experimental Study on Smartphones},'' in \emph{Proc.
  10th Symposium On Usable Privacy and Security}, 2014.

\bibitem{vatavu2015child}
R.-D. Vatavu, L.~Anthony, and Q.~Brown, ``{Child or Adult? Inferring Smartphone
  Users' Age Group from Touch Measurements Alone},'' in \emph{Proc. Conference
  on Human-Computer Interaction}, 2015.

\bibitem{Zaccagnino2021}
R.~Zaccagnino, C.~Capo, A.~Guarino, N.~Lettieri, and D.~Malandrino,
  ``{Techno-Regulation and Intelligent Safeguards},'' \emph{Multimedia Tools
  and Applications}, vol.~80, no.~10, pp. 15\,803--15\,824, 2021.

\bibitem{Tolosana2015}
R.~Tolosana, R.~Vera-Rodriguez, J.~Fierrez, and J.~Ortega-Garcia,
  ``{Feature-Based Dynamic Signature Verification Under Forensic Scenarios},''
  in \emph{Proc. 3rd International Workshop on Biometrics and Forensics}, 2015.

\bibitem{Marcos2008}
M.~Martinez-Diaz, J.~Fierrez, J.~Galbally, and J.~Ortega-Garcia, ``{Towards
  Mobile Authentication Using Dynamic Signature Verification: Useful Features
  and Performance Evaluation},'' in \emph{Proc. 19th International Conference
  on Pattern Recognition}, 2008.

\bibitem{Tolosana2015pre}
R.~Tolosana, R.~Vera-Rodriguez, J.~Ortega-Garcia, and J.~Fierrez,
  ``{Preprocessing and Feature Selection for Improved Sensor Interoperability
  in Online Biometric Signature Verification},'' \emph{IEEE Access}, vol.~3,
  pp. 478--489, 2015.

\bibitem{Chandrashekar2014}
G.~Chandrashekar and F.~Sahin, ``{A Survey on Feature Selection Methods},''
  \emph{Computers \& Electrical Engineering}, vol.~40, no.~1, pp. 16--28, 2014.

\bibitem{Saibene2023}
A.~Saibene and F.~Gasparini, ``{Genetic Algorithm for Feature Selection of EEG
  Heterogeneous Data},'' \emph{Expert Systems with Applications}, vol. 217, p.
  119488, 2023.

\bibitem{Crescenzi2019}
L.~{Crescenzi Lanna} and M.~{Grané Oro}, ``{Touch Gesture Performed by
  Children Under 3 Years Old When Drawing and Coloring on a Tablet},''
  \emph{International Journal of Human-Computer Studies}, vol. 124, pp. 1--12,
  2019.

\bibitem{mcinnes2018umap}
L.~McInnes, J.~Healy, and J.~Melville, ``{UMAP: Uniform Manifold Approximation
  and Projection for Dimension Reduction},'' 2018.

\bibitem{Acien2022}
A.~Acien, A.~Morales, R.~Vera-Rodriguez, J.~Fierrez, I.~Mondesire-Crump, and
  T.~Arroyo-Gallego, ``{Detection of Mental Fatigue in the General Population:
  Feasibility Study of Keystroke Dynamics as a Real-world Biomarker},''
  \emph{JMIR Biomed Engineering}, vol.~7, no.~2, p. e41003, 2022.

\bibitem{Gomez2023}
L.~F. Gomez, A.~Morales, J.~Fierrez, and J.~R. Orozco-Arroyave, ``{Exploring
  Facial Expressions and Action Unit Domains for Parkinson Detection},''
  \emph{PLOS ONE}, vol.~18, no.~2, pp. 1--25, 2023.

\bibitem{Daza2022}
R.~Daza, A.~Morales, R.~Tolosana, L.~F. Gomez, J.~Fierrez, and
  J.~Ortega-Garcia, ``{edBB-Demo: Biometrics and Behavior Analysis for Online
  Educational Platforms},'' \emph{arXiv}, 2022.

\end{thebibliography}
}


\begin{IEEEbiography}[{\includegraphics[width=1in,height=1.25in,clip,keepaspectratio]{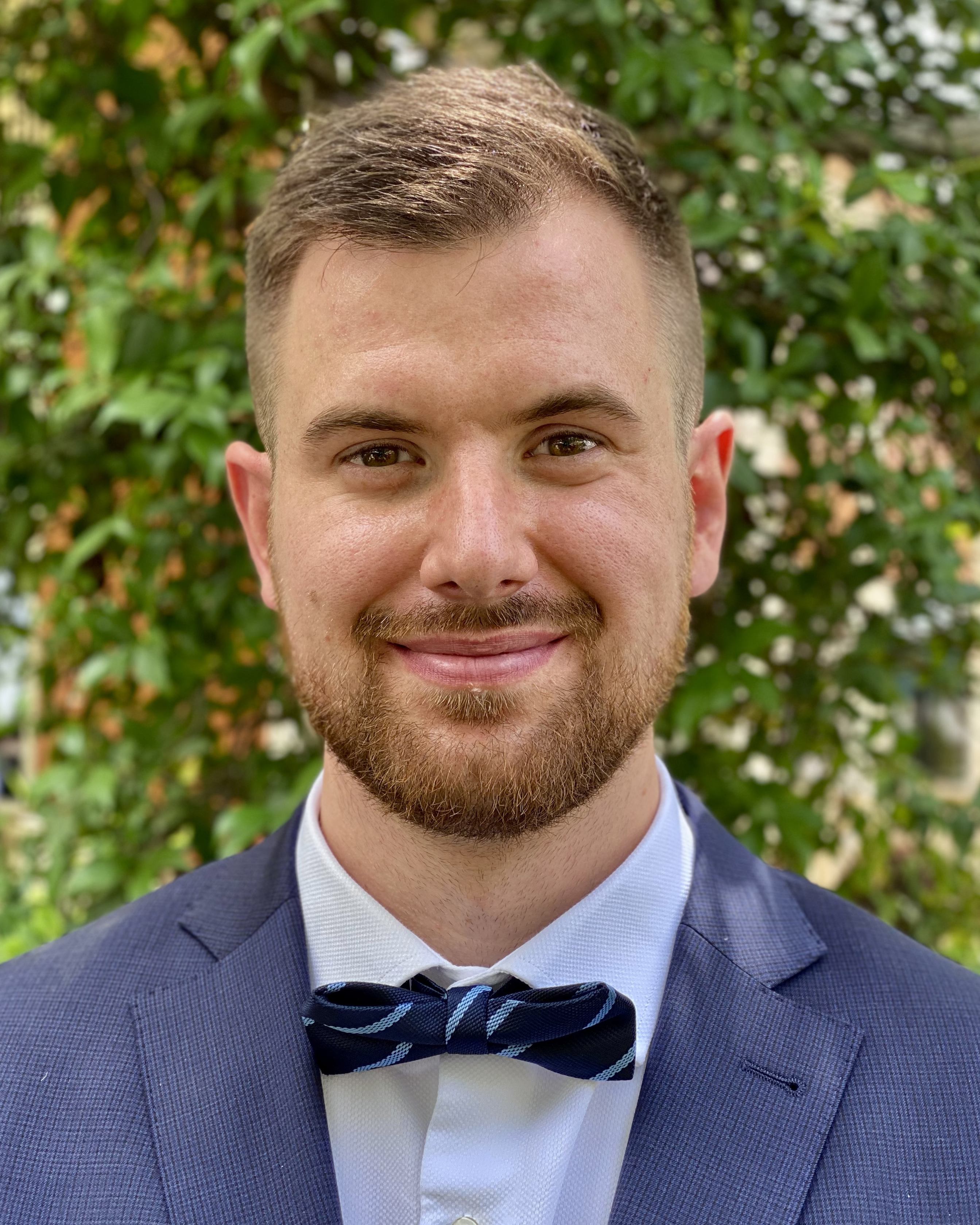}}]{Juan Carlos Ruiz-Garcia} received his B.Sc. degree in Computer Science Engineering in 2019 from the Universidad de Granada and got the M.Sc. degree in Research and Innovation in 2021 with the award of excellence from the Universidad Autonoma de Madrid, where he is currently pursuing a PhD degree in Computer and Telecommunication Engineering. In addition, in April 2020, he joined the Biometrics and Data Pattern Analytics - BiDA Lab as Pre-Doctoral Researcher at the same university. His research interests are mainly focused on the use of machine learning for e-Learning, e-Health, Human-Computer Interaction (HCI), and automatic Fall Detection Systems (FDS).
\end{IEEEbiography}

\begin{IEEEbiography}[{\includegraphics[width=1in,height=1.25in,clip,keepaspectratio]{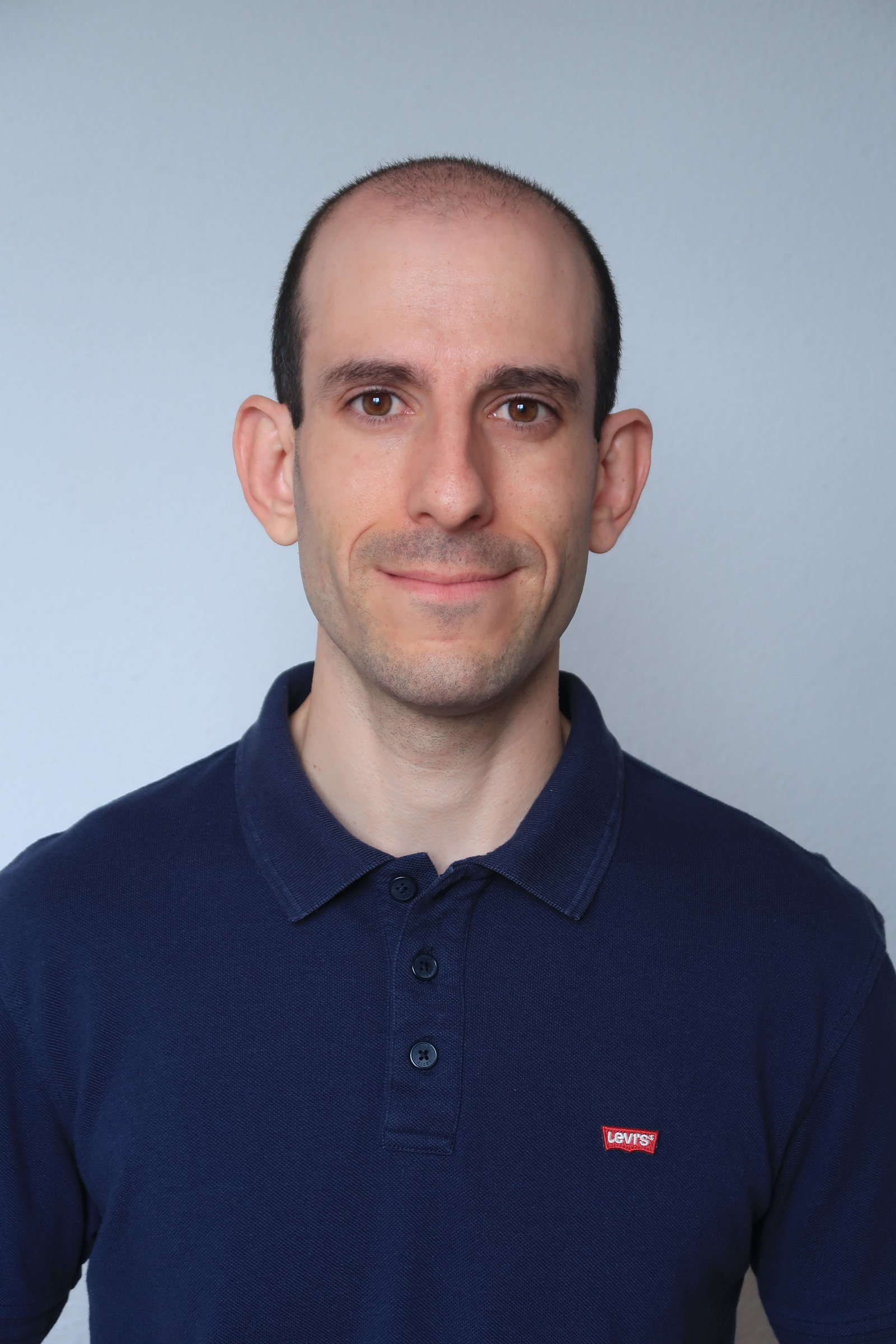}}]%
{Ruben Tolosana} received the M.Sc. degree in Telecommunication Engineering, and the Ph.D. degree in Computer and Telecommunication Engineering, from Universidad Autonoma de Madrid, in 2014 and 2019, respectively. In 2014, he joined the Biometrics and Data Pattern Analytics – BiDA Lab at the Universidad Autonoma de Madrid, where he is currently Assistant Professor. He is a member of the ELLIS Society (European Laboratory for Learning and Intelligent Systems), Technical Area Committee of EURASIP (European Association For Signal Processing), and Editorial Board of the IEEE Biometrics Council Newsletter. His research interests are mainly focused on signal and image processing, Pattern Recognition, and Machine Learning, particularly in the areas of DeepFakes, Human-Computer Interaction, Biometrics, and Health. Dr. Tolosana is actively involved in several National and European projects focused on these topics. He has also organized several workshops and challenges in top conferences such as WAMWB (MobileHCI 2023), KVC (BigData 2023), MobileB2C (IJCB 2022) and SVC-onGoing (ICDAR 2021). Dr. Tolosana has also received several awards such as the European Biometrics Industry Award (2018) from the European Association for Biometrics (EAB) and the Best Ph.D. Thesis Award in 2019-2022 from the Spanish Association for Pattern Recognition and Image Analysis (AERFAI).
\end{IEEEbiography}

\begin{IEEEbiography}[{\includegraphics[width=1in,height=1.25in,clip,keepaspectratio]{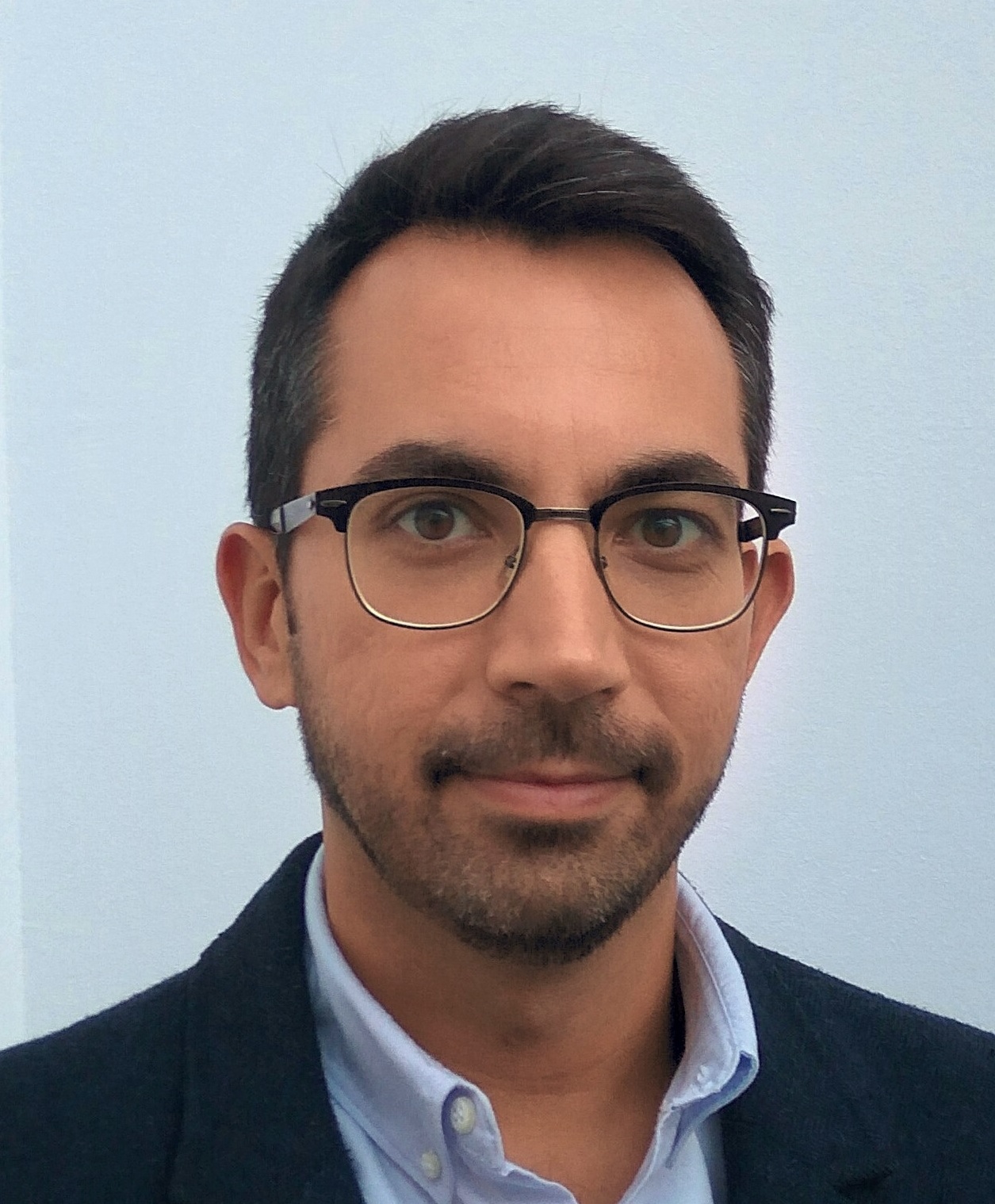}}]{Ruben Vera-Rodriguez}received the M.Sc. degree in telecommunications engineering from Universidad de Sevilla, Spain, in 2006, and the Ph.D. degree in electrical and electronic engineering from Swansea University, U.K., in 2010. Since 2010, he has been affiliated with the Biometric Recognition Group, Universidad Autonoma de Madrid, Spain, where he is currently an Associate Professor since 2018. His research interests include signal and image processing, AI fundamentals and applications, HCI, forensics, and biometrics for security and human behavior analysis. Dr. Vera-Rodriguez is actively involved in several National and European projects focused on these topics. He is author of more than 150 scientific articles published in international journals and conferences. He has served as Program Chair for some international conferences such as: IEEE ICCST 2017, CIARP 2018, ICBEA 2019 and AVSS 2022. He has also organized several workshops and challenges in top conferences such as WAMWB (MobileHCI 2023), KVC (BigData 2023), MobileB2C (IJCB 2022) and SVC-onGoing (ICDAR 2021). Ruben has received a Medal in the Young Researcher Awards 2022 by the Spanish Royal Academy of Engineering among other awards, and he is member of ELLIS Society since 2023.
\end{IEEEbiography}

\begin{IEEEbiography}[{\includegraphics[width=1in,height=1.25in,clip,keepaspectratio]{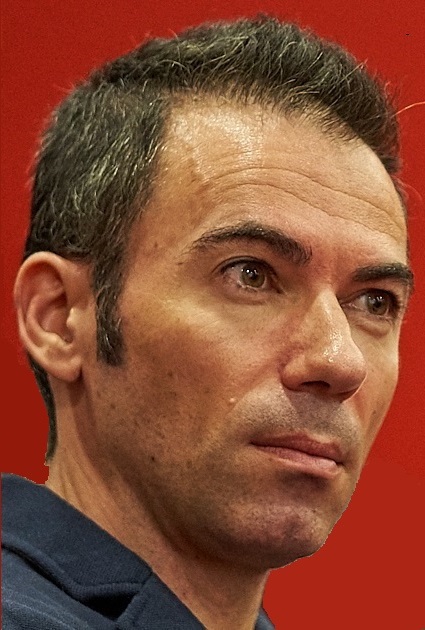}}]{Julian Fierrez} received the MSc and the PhD degrees from Universidad Politecnica de Madrid, Spain, in 2001 and 2006, respectively. Since 2004 he is at Universidad Autonoma de Madrid, where he is Associate Professor since 2010. His research is on signal and image processing, AI fundamentals and applications, HCI, forensics, and biometrics for security and human behavior analysis. He is Associate Editor for Information Fusion, IEEE Trans. on Information Forensics and Security, and IEEE Trans. on Image Processing. He has received best papers awards at AVBPA, ICB, IJCB, ICPR, ICPRS, and Pattern Recognition Letters; and several research distinctions, including: EBF European Biometric Industry Award 2006, EURASIP Best PhD Award 2012, Miguel Catalan Award to the Best Researcher under 40 in the Community of Madrid in the general area of Science and Technology, and the IAPR Young Biometrics Investigator Award 2017. Since 2020 he is member of the ELLIS Society.
\end{IEEEbiography}

\begin{IEEEbiography}[{\includegraphics[width=1in,height=1.25in,clip,keepaspectratio]{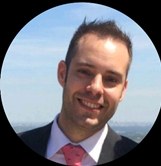}}]{Jaime Herreros-Rodriguez (JHR)} received the degree in Medicine in 2006 from Universidad Autónoma de Madrid, the title of neurologist in 2010 and he was awarded the title of Doctor in Medicine from the Universidad Complutense de Madrid (2019) with a distinction Cum Laude given unanimously for his doctoral thesis on migraine. He is also author of several publications in migraine and parkinsonism. He has collaborated with different research projects related to many neurological disorders, mainly Alzheimer and Parkinson's disease. JHR is a neurology and neurosurgery proffesor in CTO group, since 2008.
\end{IEEEbiography}

\end{document}